\documentclass[journal,onecolumn]{IEEEtran}
\usepackage[noadjust]{cite}

\ifCLASSINFOpdf
 \usepackage[pdftex]{graphicx}
\else
\fi

\usepackage{booktabs}

\ifCLASSOPTIONcompsoc
  \usepackage[caption=false,font=normalsize,labelfont=sf,textfont=sf]{subfig}
\else
  \usepackage[caption=false,font=footnotesize]{subfig}
\fi

\usepackage{hyperref}

\usepackage{siunitx}
\usepackage[capitalise]{cleveref}

\begin{document}
    %
    \title{The NIGENS General Sound Events Database}
    %
    %
    %
    
    \author{Ivo Trowitzsch, 
        Jalil Taghia,
        Youssef Kashef,
        Klaus Obermayer\\
        \textit{Neural Information Processing Group, Technische Universit\"at Berlin}
        }
    
    %
    %

    \markboth{Technical Report}%
    {Trowitzsch \MakeLowercase{\textit{et al.}}: NIGENS general sound events database}
    %



    \maketitle
    
    \begin{abstract}
        Computational auditory scene analysis is gaining interest in the last years. Trailing behind the more mature field of speech recognition, it is particularly general sound event detection that is attracting increasing attention. Crucial for training and testing reasonable models is having available enough suitable data -- until recently, general sound event databases were hardly found. We release and present a database with $714$ wav files containing isolated high quality sound events of $14$ different types, plus $303$ ``general'' wav files of anything else but these $14$ types. All sound events are strongly labeled with perceptual on- and offset times, paying attention to omitting in-between silences. The amount of isolated sound events, the quality of annotations, and the particular general sound class distinguish NIGENS from other databases.
    \end{abstract}
    

    %
    \IEEEpeerreviewmaketitle

    \section{Introduction}
    %
    %
    %
    %
    \IEEEPARstart{S}{ound}-related modeling is receiving increasing attention through the last years. Compared to speech recognition, which is more mature (and still one of the most active domains of applied machine learning research), general sound event detection (\emph{SED}) is only recently picking up pace. This is also reflected in the availability of general sound event databases, which are still scarce (\cite{DCASE2016syndata,DCASE2017challenge,urbanSounds,esc50,Fonseca2017freesound}) and have their limitations, see \cref{sec:other}. 
    
    We have built the NIGENS -- \emph{N}eural \emph{I}nformation processing group \emph{GEN}eral \emph{S}ounds -- database as the \textsc{Two!Ears} project\footnote{\url{twoears.eu}} \cite{twoears} was in need for a database of isolated high quality sound events, big enough for simulating complex acoustic scenes and the development of robust sound event detection models. To enable training of models which are able to cope with disturbances of unknown type, we included a large collection of ``general'' sounds of all kinds and sorts in addition to sounds of the detector target classes. Sounds were collected to the largest part from {StockMusic} \cite{stockmusic}, who in the meantime kindly have authorized redistribution under a non-commercial-license, such that we can now release NIGENS to the public for further research on sound-related modeling.

    \begin{table}[t]
        \caption{Sound classes}
        \label{tab:classesStats}
        \centering
        \begin{tabular}{l p{8cm} ccc}
            \toprule
            \textbf{Class} & \textbf{Description/Characteristics} & \textbf{Num files} & \textbf{Accum length} & \textbf{Avg length}\\
            \midrule
            Alarm         & Diverse sounds from old-fashioned fire bells to electronic beeps. Mostly high-pitched, discrete, sequential, very structured events; some continuous wailing. & 49 & 15m:48s & 19.4s \\
            Baby crying   & Crying babies. Mostly sequences of cries, also single sobs and squeals. Don't listen. Will break your heart.& 40 & 18m:02s & 27.1s \\
            Crash         & Crashing structures, destructive impacts; noise-like, but sudden, bursting, singular sounds. Lots of energy across wide range of frequencies. & 50 & 8m:10s & 9.8s \\
            Dog barking   & Dogs barking, mostly several times in a row. Peak of energy around 1kHz, short, discrete events. & 45 & 8m:43s & 11.6s \\
            Engine        & Long continuous sounds of running engines of different kinds, idling or changing speed. & 39 & 34m:49s & 53.6s \\
            Female scream & Short single screams of females, high-pitched, peak of energy around 1.8kHz. & 45 & 2m:46s & 3.7s \\
            Female speech & Females calmly speaking short sentences. & 100 & 4m:53s & 2.9s \\
            Fire          & Long continuous sounds of burning fires. Noise-like broadband sounds, but with higher energy in low frequencies. & 51 & 45m:20s & 53.4s \\
            Footsteps     & Diverse sounds of (individual) people walking, on all kinds of surfaces from wood to snow. Sequences of very short events. & 42 & 18m:43s & 26.8s \\
            General       & Anything outside the other classes. Discrete and continuous, single or sequential events, peaked or broadband. & 303 & 1h:31m:49s & 18.2s \\
            Knocking      & Knocking on something, mostly doors. Sequences of very short events. Most energy in low bands. & 40 & 1m:42s & 2.6s \\
            Male scream   & Short single screams of males. Peak of energy around 1.2kHz & 31 & 3m:16s & 6.4s \\
            Male speech   & Males calmly speaking short sentences. & 100 & 4m:15s & 2.6s \\
            Phone ringing & Mostly classic phones, sequences of long ringings. & 40 & 12m:16s & 18.4s \\
            Piano         & Playing piano. Both individual notes as well as monophonic sequences as well as polyphonic pieces. & 42 & 14m:33s & 20.8s \\
            \midrule
            All           &                                      & 1017 & 4h:45m:12s & 16.8s \\
            \bottomrule
        \end{tabular}
    \end{table}

    \begin{figure*}[t]
        \centering
        \subfloat[Alarm]{\includegraphics[height=0.17\textwidth]{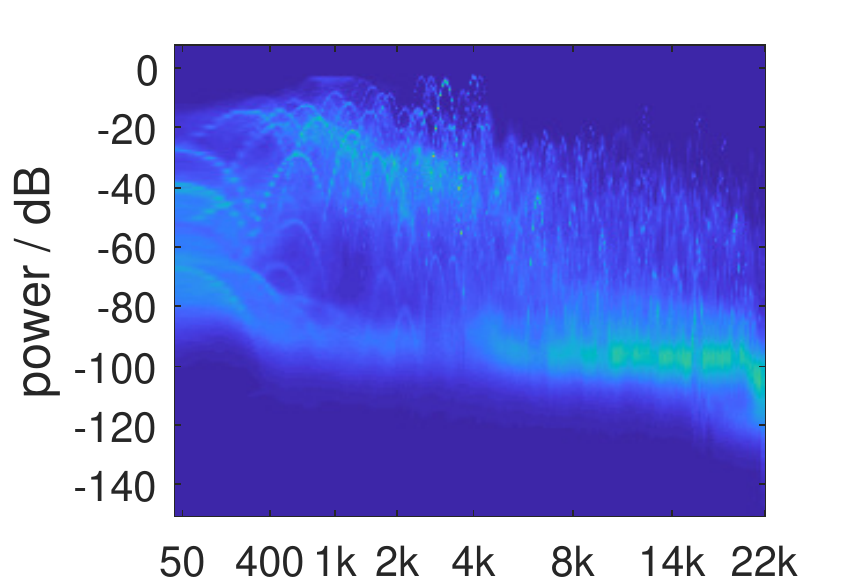}%
            \label{fig:persSpec_alarm}}
        \hfil
        \subfloat[Baby crying]{\includegraphics[height=0.17\textwidth]{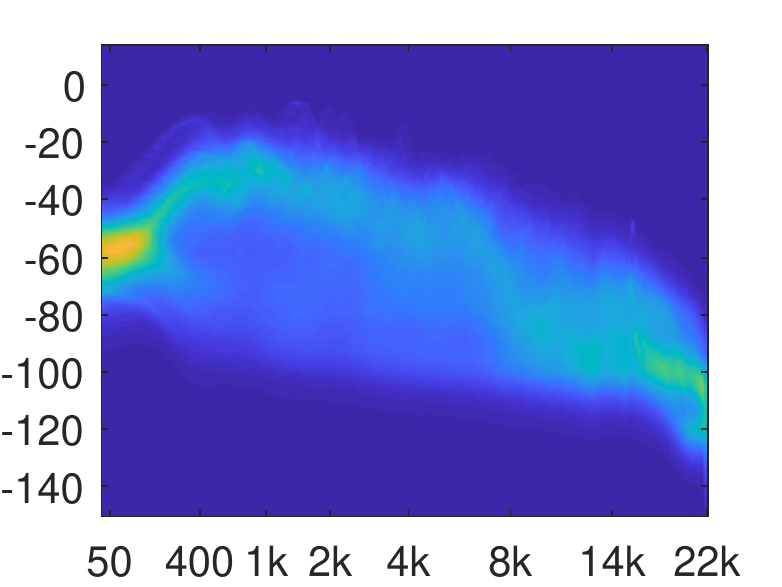}%
            \label{fig:persSpec_baby}}
        \hfil
        \subfloat[Female speech]{\includegraphics[height=0.17\textwidth]{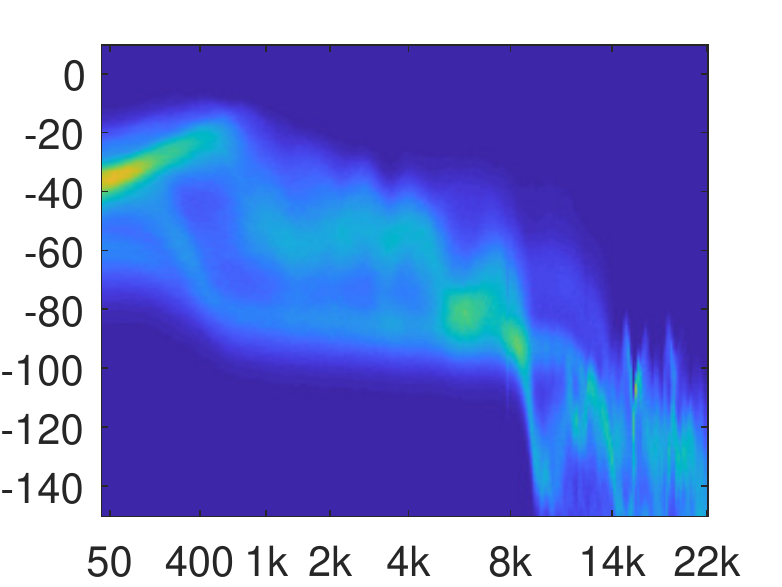}%
            \label{fig:persSpec_fspeech}}
        \hfil
        \subfloat[Crash]{\includegraphics[height=0.17\textwidth]{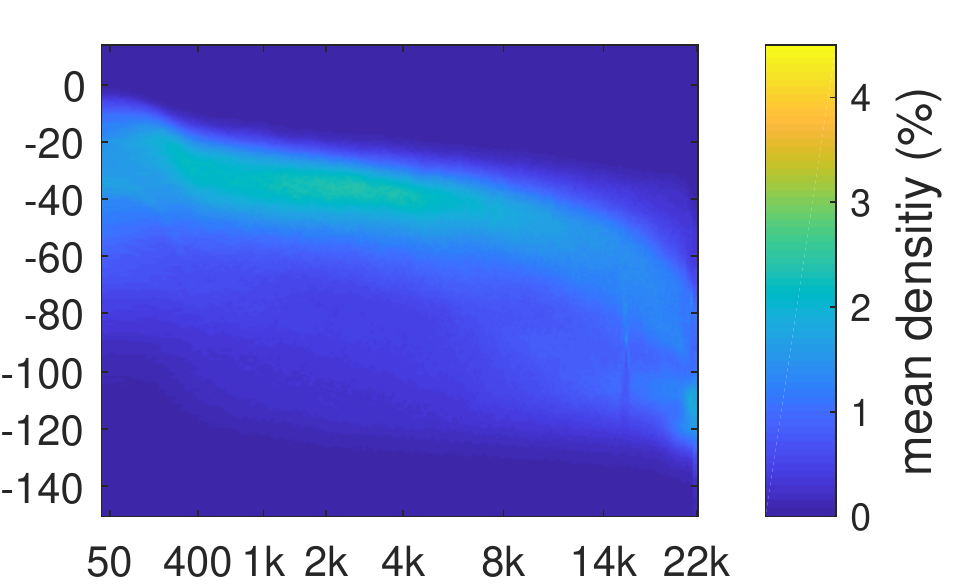}%
            \label{fig:persSpec_crash}}
        \\
        \subfloat[Phone ringing]{\includegraphics[height=0.17\textwidth]{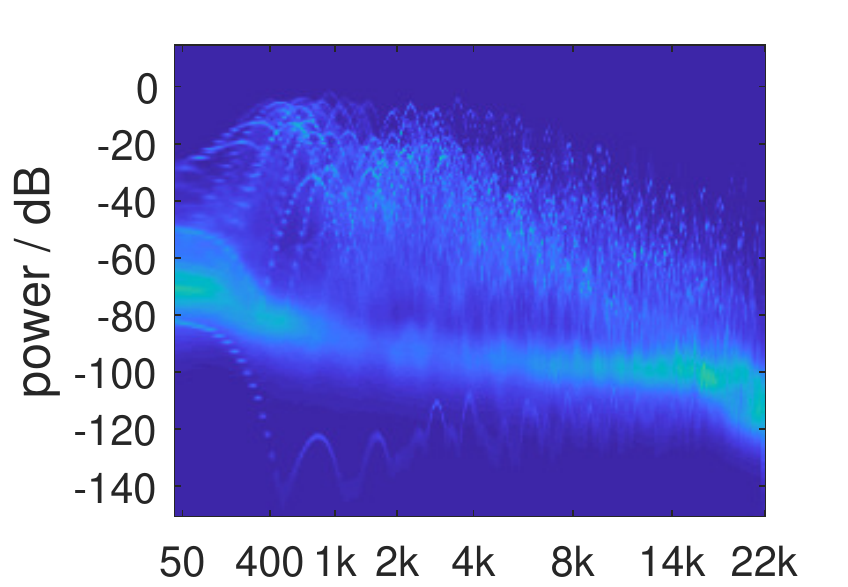}%
            \label{fig:persSpec_phone}}
        \hfil
        \subfloat[Female scream]{\includegraphics[height=0.17\textwidth]{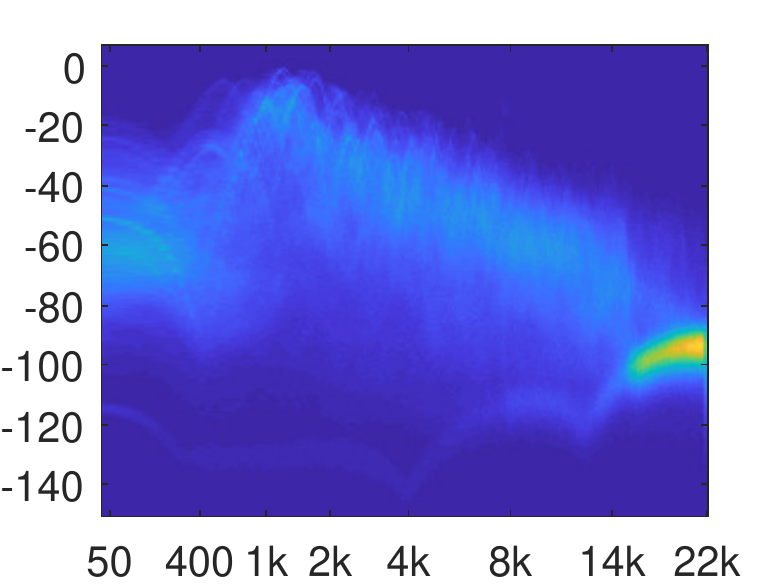}%
            \label{fig:persSpec_fscream}}
        \hfil
        \subfloat[Male speech]{\includegraphics[height=0.17\textwidth]{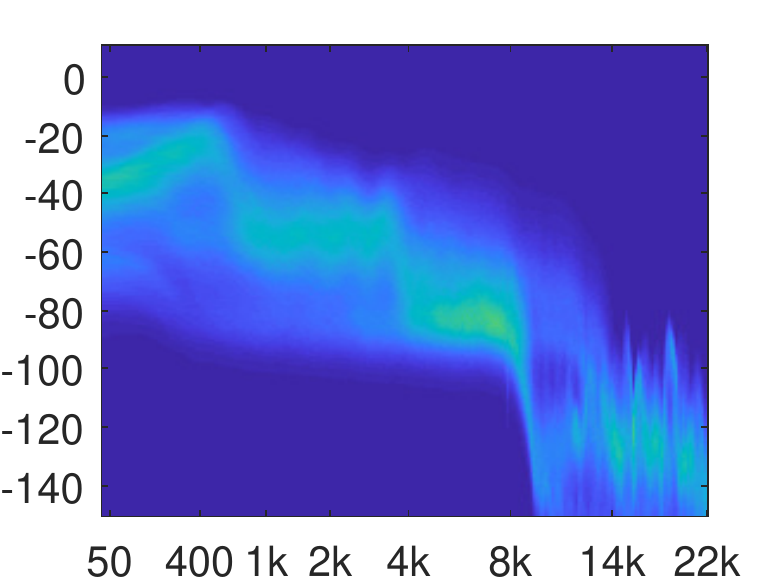}%
            \label{fig:persSpec_mspeech}} 
        \hfil
        \subfloat[Engine]{\includegraphics[height=0.17\textwidth]{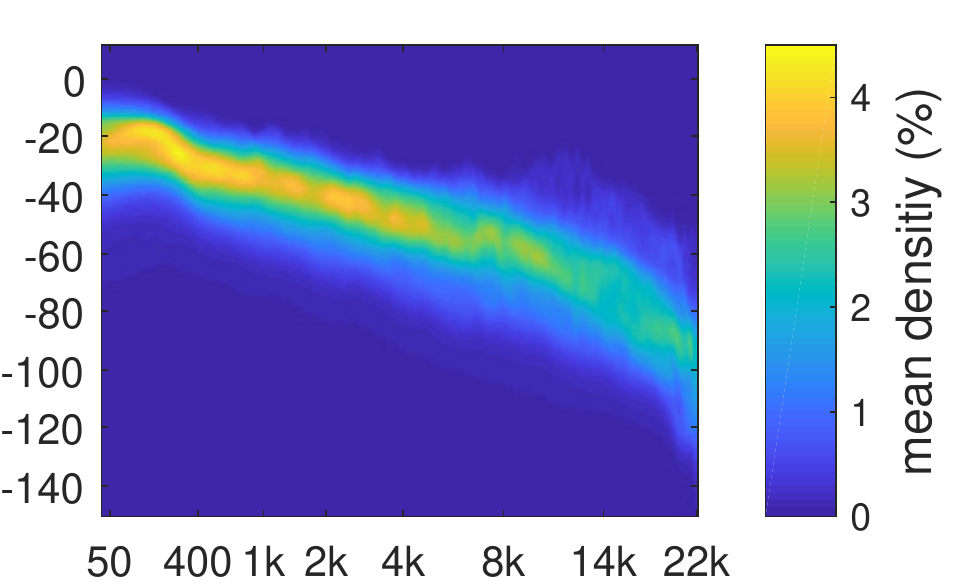}%
            \label{fig:persSpec_engine}}
        \\
        \subfloat[Piano]{\includegraphics[height=0.17\textwidth]{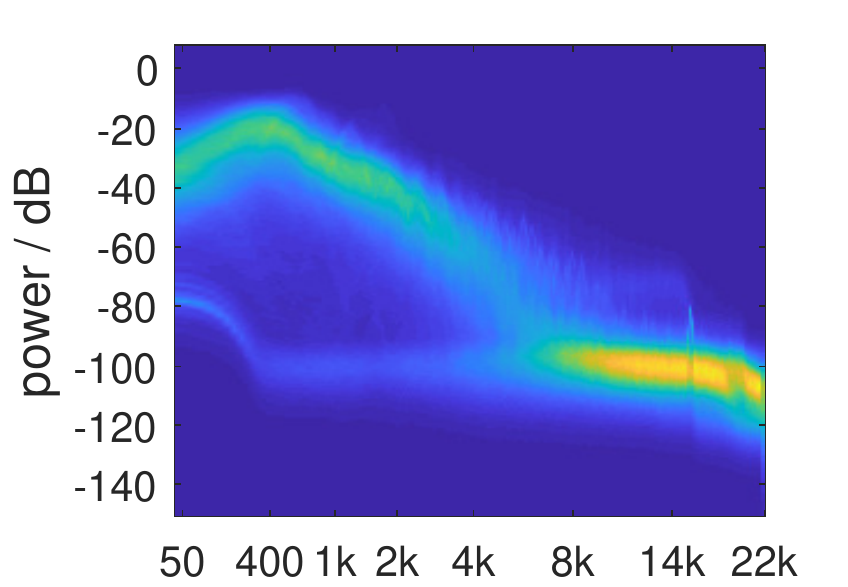}%
            \label{fig:persSpec_piano}}
        \hfil
        \subfloat[Male scream]{\includegraphics[height=0.17\textwidth]{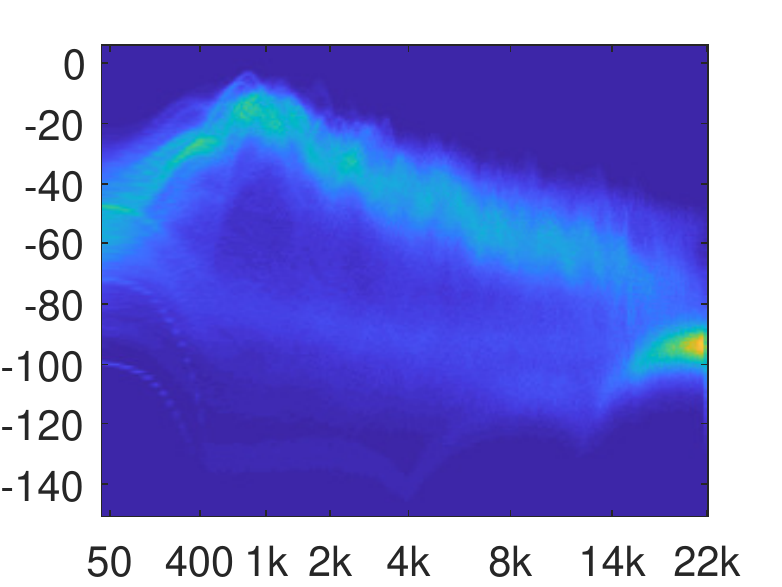}%
            \label{fig:persSpec_mscream}} 
        \hfil
        \subfloat[Footsteps]{\includegraphics[height=0.17\textwidth]{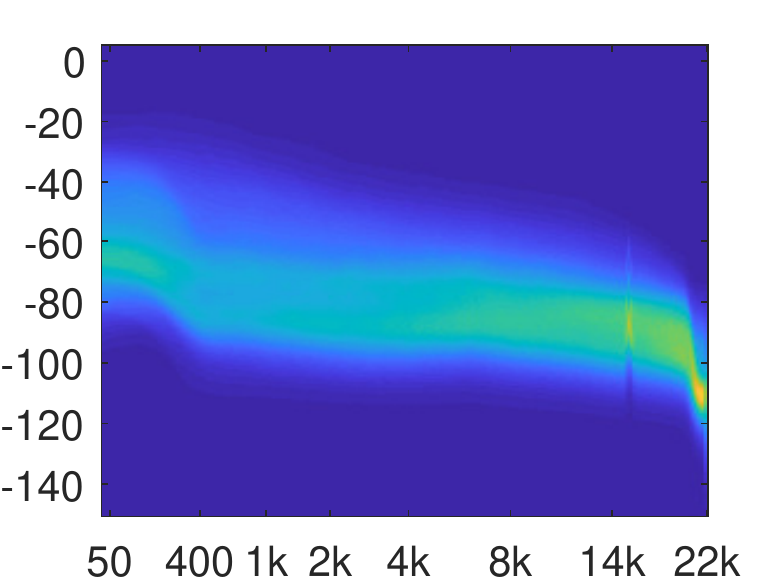}%
            \label{fig:persSpec_footsteps}}
        \hfil
        \subfloat[Fire]{\includegraphics[height=0.17\textwidth]{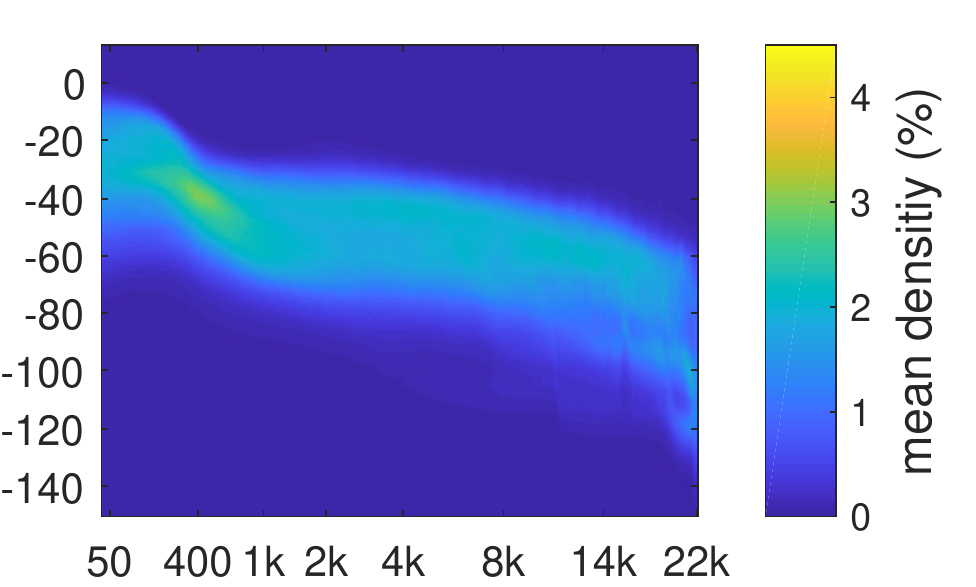}%
            \label{fig:persSpec_fire}} 
        \\
        \hspace{0.23\textwidth}
        \hfil
        \subfloat[Dog barking]{\includegraphics[height=0.19\textwidth]{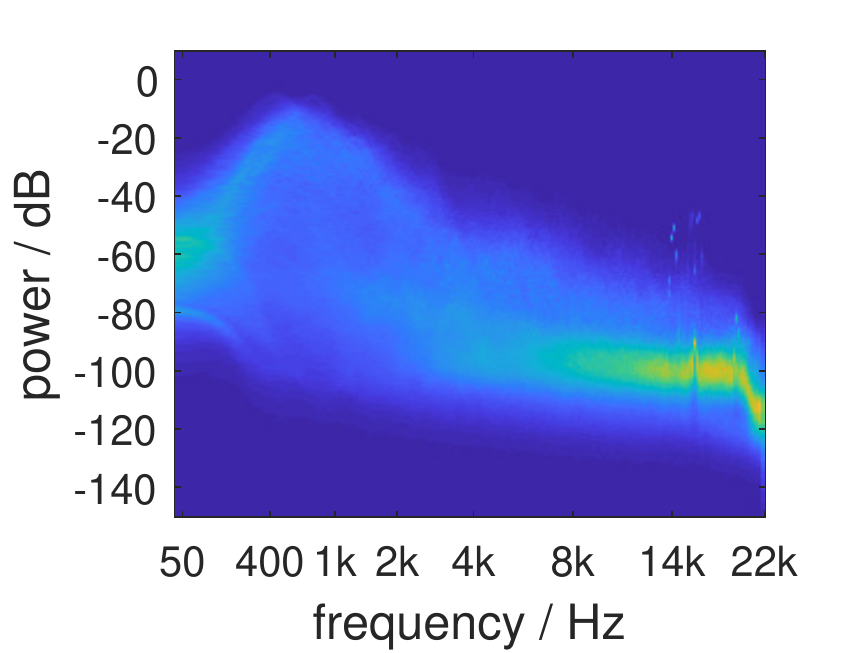}%
            \label{fig:persSpec_dog}} 
        \hfil
        \subfloat[Knocking]{\includegraphics[height=0.19\textwidth]{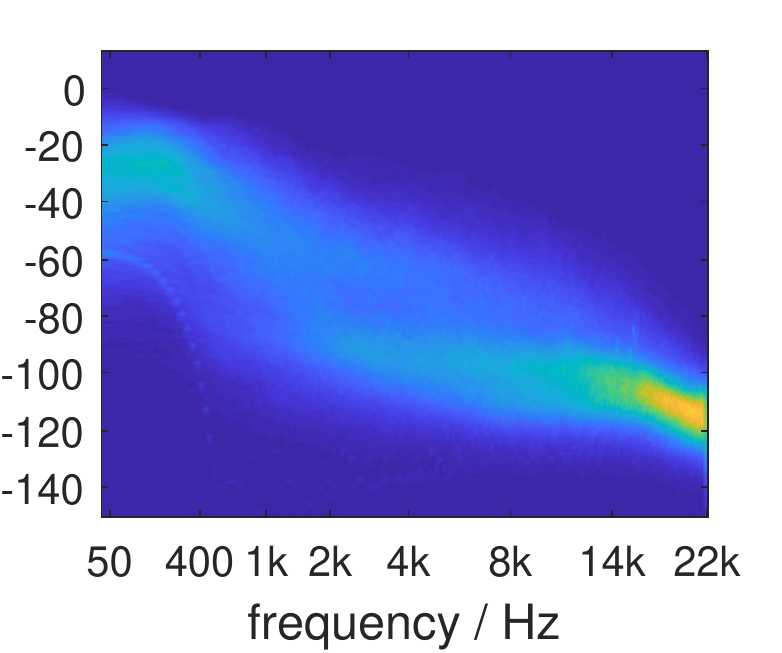}%
            \label{fig:persSpec_knock}}
        \hfil
        \subfloat[General]{\includegraphics[height=0.19\textwidth]{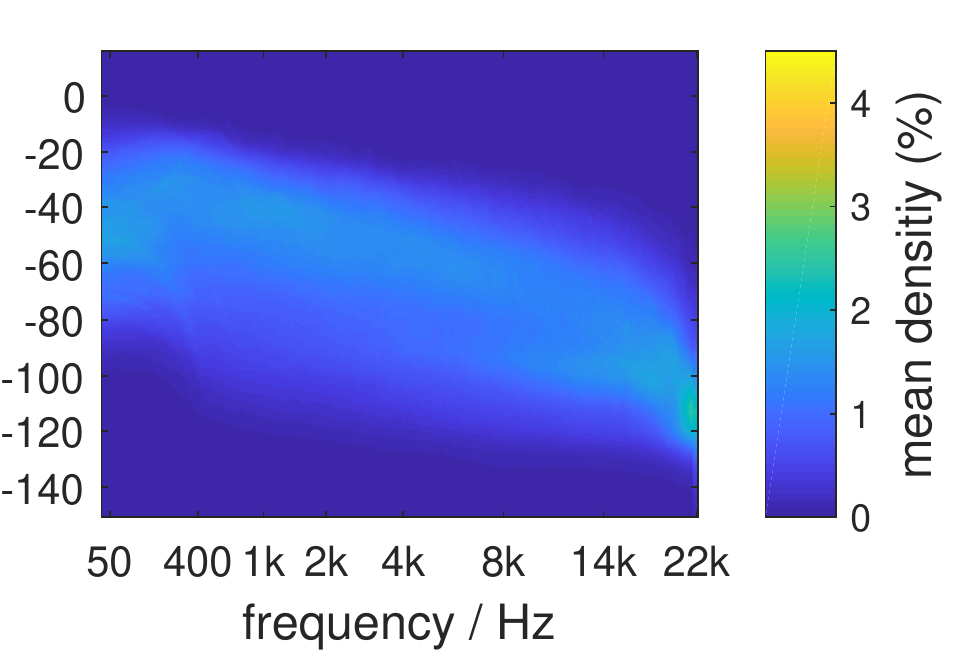}%
            \label{fig:persSpec_general}}
        \caption[NIGENS class-average persistence spectra]{Class-average persistence spectra. Temporal ``density'' of sounds over frequency and power is displayed, showing power distribution over frequencies, but also sound structure. Density scales equal for all plots. }
        \label{fig:class_persSpecs}
    \end{figure*}

    \section{NIGENS Contents}
    NIGENS is publicly accessible at \cite{nigens}. It consists of \num{1017} audio files of various lengths (between \SI{1}{\second} and \SI{5}{\minute}), in total comprising \SI{4}{\hour}:\SI{45}{\minute}:\SI{12}{\second} of sound material. Mostly, sounds are provided with $32$-bit precision and \SI{44100}{\hertz} sampling rate. Files contain isolated sound events, that is, without superposition of ambient or other foreground sources.
    
    Fourteen distinct sound classes are included: \texttt{alarm}, \texttt{crying baby}, \texttt{crash}, \texttt{barking dog}, \texttt{running engine}, \texttt{burning fire}, \texttt{footsteps}, \texttt{knocking on door}, \texttt{female} and \texttt{male speech}, \texttt{female} and \texttt{male scream}, \texttt{ring\-ing phone}, \texttt{piano}.
    Additionally, there is the \texttt{general} (“anything else”) class. 
    \cref{tab:classesStats} provides descriptions and statistics for all classes. \cref{fig:class_persSpecs} shows \emph{persistence spectra} for all classes, averaged over all sounds of each class. Persistence spectra display temporal ``density'' (rate of occurrence), over frequency and power. They are computed based on short-time Fast-Fourier Transform (FFT) spectrograms, but in contrast to them can be reasonably averaged. Compared to pure power spectra, they are able to also depict structure of sounds. Note, for instance, the similar structure of alarm and phone, versus the similar structures of engine and fire. The \texttt{general} (\cref{fig:persSpec_general}) class shows, as expected, a broadband, unstructured spectrum, since there are so many different types of sounds included. It is described in the section below.
    
    \begin{figure}[t]
        \centering
        \includegraphics[width=0.7\textwidth]{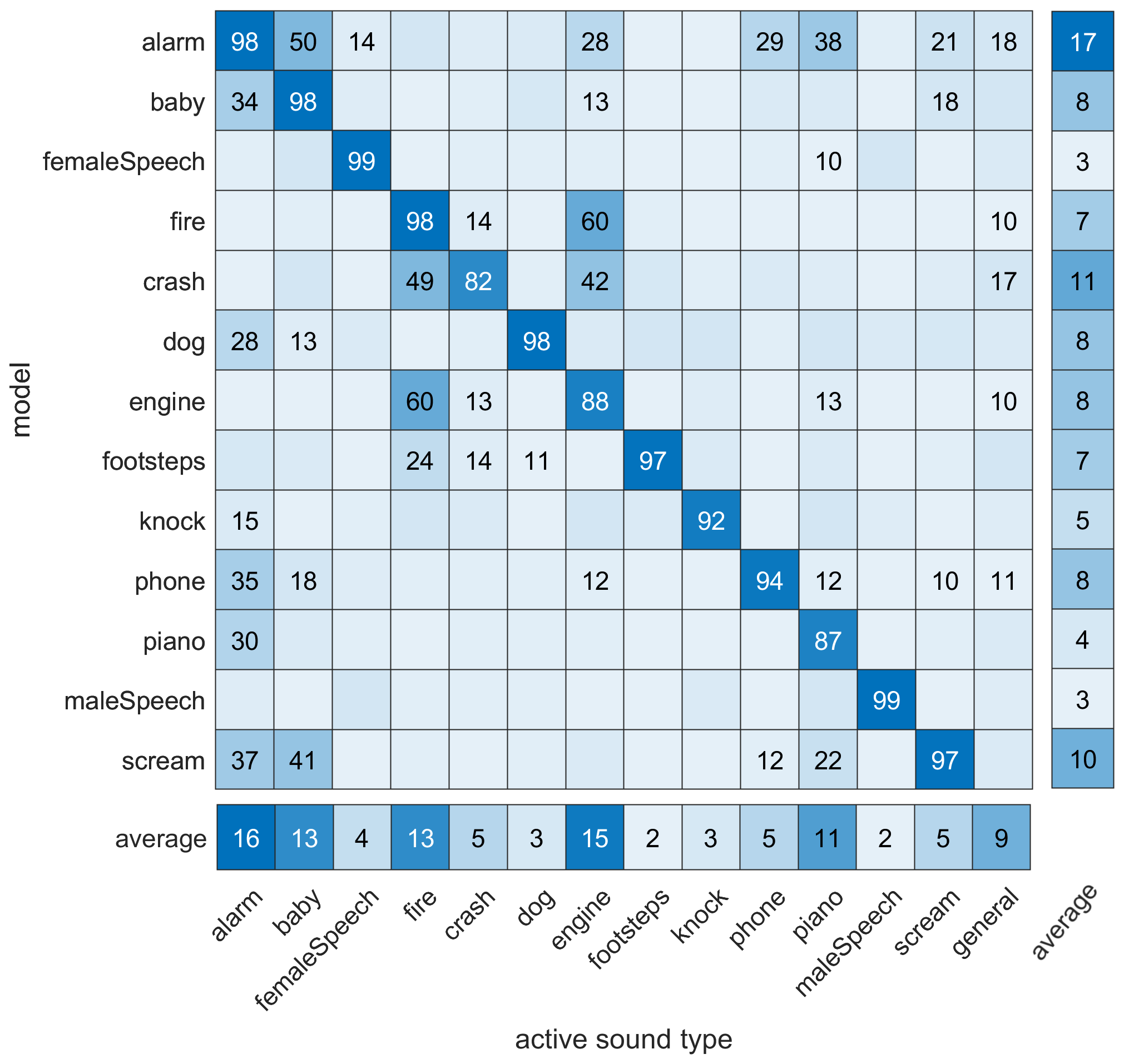}
        \caption[NIGENS confusion matrix]{NIGENS confusion matrix, values are classification rates in percent in scenes with one source. Rows correspond to trained SED models (models described in \cite{trowitzsch2019}), columns to types of active sound events. For better readability, only values of at least \SI{10}{\percent} are displayed with label. The averages are without sensitivities (the values for matching sound and model type), that is, they are misclassification averages.}
        \label{fig:nigens_confusionMatrix}
    \end{figure}
    
    \cref{fig:nigens_confusionMatrix}, without elaboration on the models producing these results (confer \cite{trowitzsch2019} if interested), shows a classification rate confusion matrix of the NIGENS types and corresponding trained models. The values are percentages of (\SI{500}{\milli\second}-) segments of binaural auditory one-source-scenes classified by the models as their corresponding type. On the diagonal are the sensitivities (positive classification rates, or detection rates), all other values are misclassifications: for example, the ``femaleSpeech'' model classified (correctly) \SI{99}{\percent} of the actual femaleSpeech segments as femaleSpeech, but also (wrongly) \SI{10}{\percent} of the ``piano'' segments. These classification rates, although of course specific to the models which produced them, can serve as an indicator of the overlap of the sound classes.

    Sounds for all but the \texttt{speech} and \texttt{scream} (and few \texttt{general}) classes were attained from {StockMusic} \cite{stockmusic}, the online redistributor of the Sound Ideas sounds library, offering individual files rather than pre-compiled collections. \texttt{Scream} and the remaining \texttt{general} sounds were collected from \href{https://freesound.org}{Freesound.org} \cite{freesound}.     
       
    \subsection*{``General'' Sound Class}
    Often, sound event detectors are trained to discriminate between a target class and all other target classes, sometimes added by broadband-like ambient noise. If testing is done the same way, this certainly produces highest performances. However, this approach lacks a real-world circumstance: there will always be a lot of sound events occurring that were not part of any target training class, many of them discrete and not noise-like. \cite{Mesaros2018_TASLP} also identify this as a key difference between the DCASE 2016 SED synthetic and real audio tasks. To explicitly take this into account, and help better define target detector models against sounds different from other target classes, the \texttt{general} class was collected. This class contains sounds intended both as ``disturbance'' sound events (superposing) and as counterexamples to the target sound classes.
    
    The \texttt{general} class is a pool of sound events \emph{other} than the $ 14 $ distinguished target sound classes, containing as heterogeneous sounds as possible. For example, it includes nature sounds such as wind, rain, or animals, sounds from human-made environments such as honks, doors, or guns, as well as human sounds like coughs.

    \subsection*{Speech sound files}
    The database contains female and male speech sounds, which were compiled from the GRID \cite{grid} and TIMIT \cite{timit} copora. The latter one unfortunately is attached with a restrictive license, preventing us from redistributing the respective files. The associated event on- and offset time annotations created by us, however, are made available by us; and in \cite{nigens}, we list the TIMIT files used, leaving anybody with the possibility to get hold of these files by themselves.
    
    \subsection*{Event On- and Offset Times}
    In order to effectively train models that detect sound events of particular classes, sounds have been annotated by time stamps indicating perceptual on- and offsets of occurring sound events. Wave files are thus accompanied by an annotation (.txt) file that includes on- and offset times of that file's sound events. The \texttt{general} sounds do not come with on- and offset time annotations, since these files do not constitute any coherent sound class (to the contrary, by design) and are not intended to be positive examples for classifiers.
    
    In contrast to other SED data sets, only active sound were labeled as actual sound events, that is, times of silence are not part of sound events with this labeling. Positive labeling across ``gaps'' in sound events is more of a semantic-logical labeling (referring to a series of individual phone rings as ``phone ringing'', for example), but it can be assumed that this complicates training since the direct correlation of physical features and label gets lost.
    
    \pagebreak
    
    Since each sound file may contain an arbitrary number of instances of the same event class, the list of on- and offset times (in seconds) for each sound file is stored in a separate text file:
    
    {\small
    \begin{verbatim}
    0.002268 1.814059
    6.061224 7.922902
    12.176871 13.827664
    ...
    \end{verbatim}
    }

    \subsection*{File Lists}
    For easy parsing and better comparability of results based on this data set, we provide the following file lists: \verb|all.flist| is a text file containing a list of all included sounds. \verb|NIGENS_8-foldSplit_fold<x>.flist| are lists of eight disjunct folds of equal size. If you do only one train-test-set split for model training and evaluation of generalization performance, we suggest using fold 1-6 for training (and do cross-validation on them, if applicable) and 7-8 for testing. Using the same train-test-set splits across works will increase comparability.
    
    The same file lists are provided as versions without TIMIT files included, in case you don't have access to these.

    \section{License}
    You are free to use this database non-commercially under \href{https://creativecommons.org/licenses/by-nc-nd/4.0/legalcode}{Creative Commons Attribution-NonCommercial-NoDerivatives 4.0} license.

    \section{Using NIGENS}
    If you use this data set, please cite as:\\
    \\
    Ivo Trowitzsch, Jalil Taghia, Youssef Kashef, and Klaus Obermayer (2019). \emph{The NIGENS general sound events database}. Technische Universit\"at Berlin, Tech. Rep. arXiv:1902.08314 [cs.SD]\\

    In \cite{trowitzsch2017}, we have developed and analyzed a robust binaural sound event detection training scheme on NIGENS sounds. In \cite{trowitzsch2019}, this work was extended in order to join sound event detection and localization. Both works were done utilizing the \emph{Auditory Machine Learning Training and Testing Pipeline} \cite{amlttp}, AMLTTP, which can process the file lists, on- and offset time annotation files and sounds provided by NIGENS out of the box. AMLTTP is particularly suited for sound event detection model training; among other features, it enables straight-forward generation of complex spatial polyphonic sound scenes together with polyphonic annotations, from databases with isolated sounds like NIGENS.
    
    \section{Other Data Sets} \label{sec:other}
    In the following, we list other datasets we are aware of that contain more or less \emph{isolated} sound events, which enables generation of well-defined acoustic scenes of specific complexity.
    \begin{itemize}
    \item 
    The DCASE 2016 \cite{DCASE2016syndata} task 2 (synthetic audio sound event detection) data set consists of $ 20 $ short mono sound files for each of $ 11 $ sound classes (from office environments, like \texttt{clearthroat}, \texttt{drawer}, or \texttt{keyboard}), each file containing one sound event instance. Sound files are annotated with event on- and offset times, however silences between actual physical sounds (like with a phone ringing) are not marked and hence ``included'' in the event. This data set is very small.
    \item 
    The DCASE 2017 \cite{DCASE2017challenge} rare sound events task data set contains isolated sound events for three classes: $ 148 $ crying babies (mean duration \SI{2.25}{\second}), $ 139 $ glasses breaking (mean duration \SI{1.16}{\second}), and $ 187 $ gun shots (mean duration \SI{1.32}{\second}). As with the DCASE 2016 data, silences are not excluded from active event markings in the annotations. While this data set contains many samples per class, there are only three classes, which limits possible scene generation and also generalization of obtained results considerably. 
    \item 
    The UrbanSound and UrbanSound8k datasets \cite{urbanSounds} provide a large database with \num{1302} different sound files (containing \SI{27}{\hour} of audio) distributed across ten classes of urban environments, like car horn, dog bark, or jackhammer. Sounds originated from \url{Freesound.org} and were enhanced by manual annotations of sound event starting and ending times. Unfortunately, sound events are not necessarily isolated, but instead marked with saliency annotations whether the respective event is perceived to be in the foreground or background. Using the UrbanSound8k dataset, which is a subset of UrbanSound with slices of \SI{4}{\second} length, and constraining to foreground instances, could be a way to at least obtain events that are perceived dominant.
    \item 
    The ESC-50 dataset \cite{esc50} comprises \num{2000} \SI{5}{\second}-clips of $ 50 $ different classes across natural, human and domestic sounds, again, drawn from Freesound.org. While it has been attempted to extract sounds restricted to foreground events with limited background noise, events are not truly isolated. Also, events are not annotated with event on- and offset times.
    \item 
    The Freesound Datasets \cite{Fonseca2017freesound} consist of audio samples from \url{Freesound.org}, organized in a hierarchy based on the AudioSet Ontology, with verified event labels. It is an ongoing project (albeit a very large one already) more than a completed dataset, aiming to increase the number of audio files with (through crowd-sourcing) verified labels. However, these annotations are weak labels, as they only provide information about the existence of a sound event throughout the file, but no information about when it occurs. As with UrbanSound, events are not necessarily isolated, but are labeled to be predominant or not. As of presented in \cite{Fonseca2017freesound}, \num{20206} audio clips (\SI{92.5}{\hour}) are already labeled and verified with predominant events.
    \end{itemize}

    \section{Attribution}
    The largest part of the sounds was acquired from and kindly granted redistribution for research under above license by \href{https://stockmusic.sourceaudio.com}{StockMusic.com} \cite{stockmusic}. These files have been watermarked to enable misuse detection. Please comply with the license.
    Speech sounds were compiled from the GRID \cite{grid} and TIMIT \cite{timit} copora. 
    Several sounds were downloaded and included from \href{freesound.org}{https://freesound.org} \cite{freesound} under attribution licenses, a list can be found in \cite{nigens}.

    \section{Funding}
    The collection of this dataset was partly funded from the European Union’s Seventh Framework Programme for research, technological development and demonstration under grant agreement no 618075.

    
    %

    \appendices
%

%
%
%
    
    \ifCLASSOPTIONcaptionsoff
    \newpage
    \fi

\end{document}